\documentclass[preprint2]{aastex}
\newcommand{\header}[1]{\multicolumn{1}{c}{\textrm{#1}}}
\newcommand{\Teff}{$T_{eff}$}
\newcommand{\lgg}{$\log\,g$}

\shorttitle{Wavelength calibration of the Hamilton echelle spectrograph}
\shortauthors{Pakhomov et al.}

\begin{document}

\title{Wavelength calibration of the Hamilton echelle spectrograph}

\author{Yu.Pakhomov}
\affil{Institute of Astronomy, Russian Academy of Sciences, Moscow, Russia}
\email{pakhomov@inasan.ru}
\and
\author{G. Zhao}
\affil{National Astronomical Observatories of China, Chinese Academy of
Sciences, Beijing, China}

\begin{abstract}
We present the wavelength calibration of the Hamilton echelle spectrograph
(the Lick observatory). The main problem of the calibration arises from the fact
that thorium lines are absent in the spectrum of ``ThAr'' hollow-cathode lamp
now under the operation. On the other hand, numerous unknown strong lines are
present in the spectrum. These lines was identified with titanium. We estimate
the temperature of the lamp gas which permits us to calculate the intensities of
the lines, and to select a large number of relevant \ion{Ti}{1} and \ion{Ti}{2}
lines. The titanium line list for the Lick hallow-cathode lamp is presented. The
wavelength calibration using this line list was made with accuracy about
0.006~\AA.
\end{abstract}


\section{Introduction}

Hallow-cathode lamps produce a great number of lines in the optical band which
make them indispensable for the accurate wavelength calibration of echelle
spectra. The essential prerequisite, however, is the knowledge of wavelengths of
reference spectral lines with the accuracy better than the resolution,
$\delta\lambda \sim0.1\lambda/R$. Usually, the same line list for the particular
type of a lamp is used in data reduction of spectral observations.

On September 1, 2012, spectra of the 18 stars were taken on the Shane 3-m
telescope of the Lick observatory using the Hamilton echelle spectrograph
(title of the proposal ``A systematic study of NLTE abundance of nearby
dwarfs'', PI: Zhao G., NAOC, China). The resolving power of $R\approx60\,000$
and CCD {\it e2v CCD203-82} (4k x 4k, 12$\mu$, Dewar \#4) were used for this set
of observations. The preliminary processing of the images was made using the
MIDAS package. We extract 115 echelle orders with absolute numbers between 56
and 170 that corresponds to the wavelength region of 3350--10200\AA. 

The crucial stage of the processing is the wavelength calibration. This issue is
of special importance for spectral observations of stars with exoplanets, for
which the Hamilton spectrograph is actively used. The wavelength standards is
performed by means of the hallow-cathode lamp manufactured by S\&J Juniper \& Co
(Serial No. 531495). The lamp with the designation LICK-HCL-002 presumably has
the thorium cathode, quartz envelope, and filled with a mixture of argon (90\%)
and neon (10\%). In our case, several spectral lines of Ar have been identified
to launch the calibration with MIDAS command \emph{IDEN/ECH} using the standard
line list of ThAr lamp from file \emph{thar100.tbl}. However, this procedure has
identified only $\sim5$\% of all detected lines (63 lines from 1326) using 2D
solution of standard echelle relation, far below what is needed to perform
accurate calibration. Moreover, only 36 orders from 115 has identified
lines. In case of independently calibration for each order in MIDAS, the number
of identified lines $\sim18$\%, and typical error of automatic calibration is
0.1-0.7\AA. Only about 20 orders has accuracy about 0.01\AA, but most of they
disposed at region of $\lambda<5000\AA$ with low signal to noise ratio. A
similar situation has arisen in the case of the ThAr more detailed line list
from NOAO \citep{NOAO_ThAr}. Thus, the automatic calibration is wrong due to the
mismatch of the lamp spectrum and the line list of ThAr.
On the other hand, the calibration may be done using only argon lines in
manual mode. For this, we would need at least a few evenly distributed lines
in each order and that is available only for about a third of the orders,
therefore making the calibration impossible.

Given this unexpected failure we scrutinize the problem and come up with the 
solution which is a major subject of the present communication. The results
hopefully will be of interest for observers who face the need for the wavelength
calibration with this instrument. We substantially expand the line list of argon
appropriate for the wavelength calibration thanks to a great number of titanium
lines we identified in the lamp spectrum.

The paper content is as follows. We first analyse the spectrum of the lamp
LICK-HCL-002 and identify \ion{Ti}{1} lines (Section 2).  We then determine the
gas temperature to compile the line list of Ti appropriate for the spectrum
produced by the lamp (Section 3). In Section 4, we use the new line list for the
wavelength calibration of the Hamilton spectrograph. 

\section{Lamp spectrum identification}

\begin{figure*}
\centering
\resizebox{0.85\hsize}{!}{\includegraphics[clip]{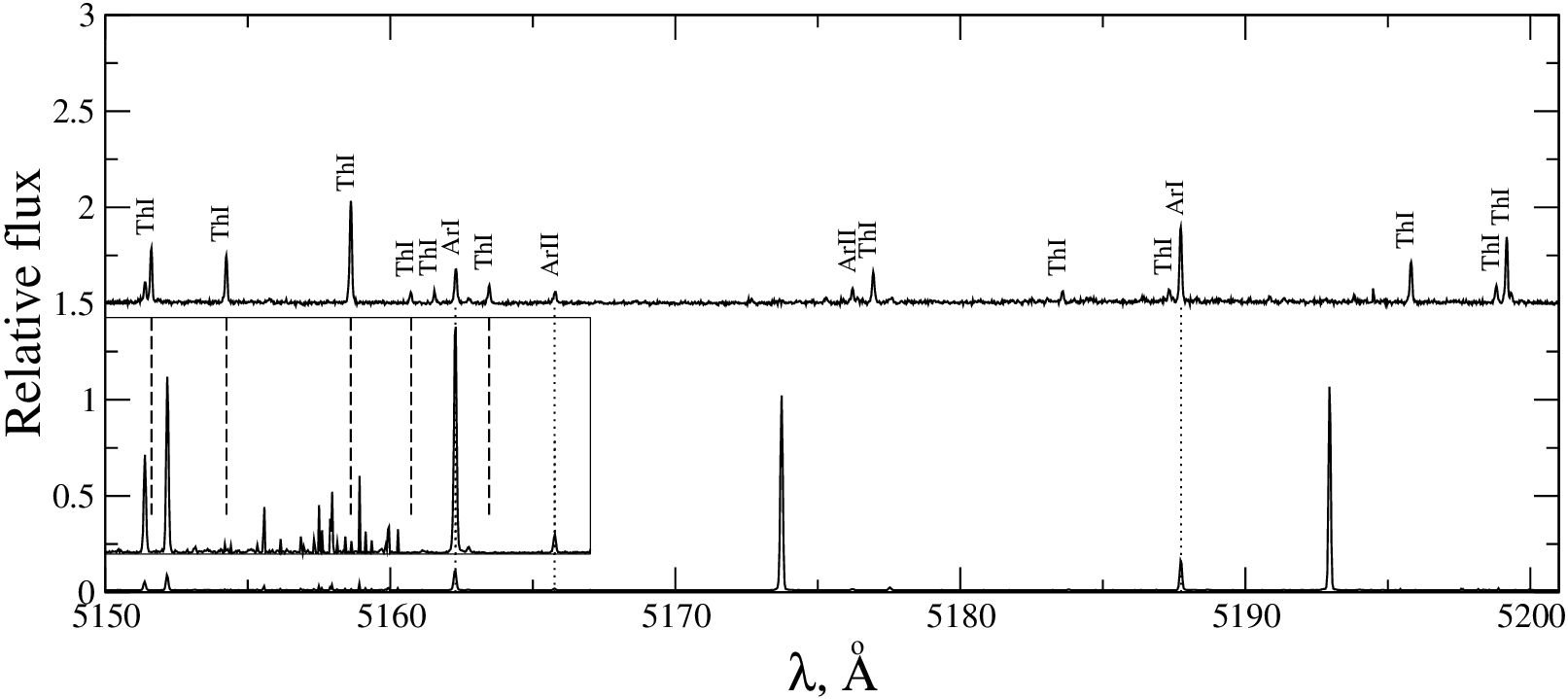}}
\figcaption{Spectrum of the Lick lamp from the 110-th order (bottom) and 
NTT spectrum of ThAr lamp (top). The dotted line indicate common identified Ar
lines. Inset shows 10-fold intensity zoom of the Lick lamp spectrum with 
dashed lines emphasising the absence of thorium lines.}
\label{fig:cmp1}
\end{figure*}

\begin{figure*}
\centering
\resizebox{0.85\hsize}{!}{\includegraphics[clip]{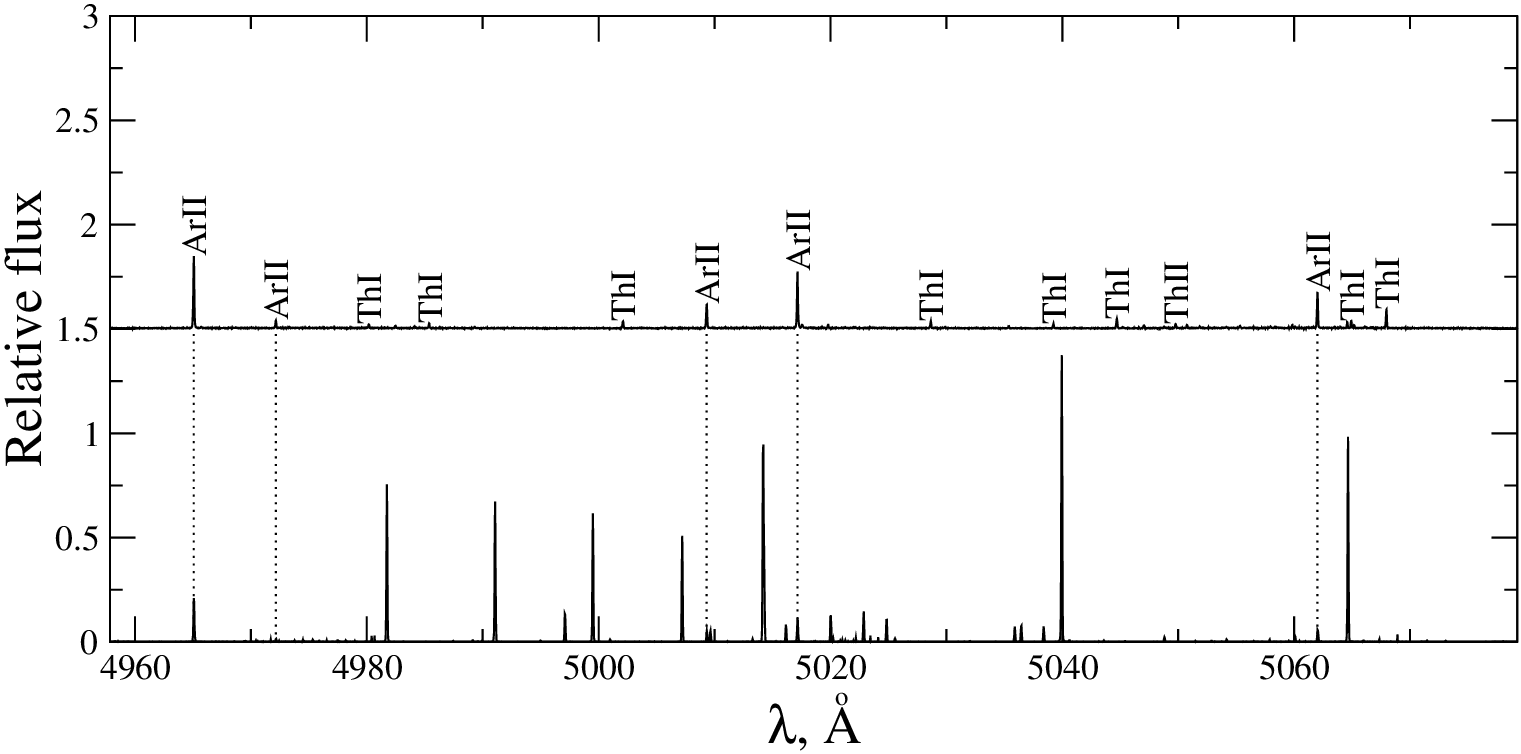}}
\figcaption{The same as Fig. 1 but for the 114-th order.}
\label{fig:cmp2}
\end{figure*}

The relative intensity of spectral lines in the spectrum of hallow-cathode lamps
depends generally on a number of factors including the current, voltage, density
of filling gas, service time, and possibly some other factors. The direct
outcome of this is that one never faces exactly the same spectra for the
arbitrary pair of lamps. Yet one always can find a set of common lines in
spectra of different lamps. In Fig.~\ref{fig:cmp1}, we show the spectrum of the
110-th echelle order compared with the spectrum of ThAr lamp obtained with the
EMMI REMD echelle spectrograph installed on the ESO NTT telescope. This lamp was
used earlier e.g., in the paper \citep{2012MNRAS.424.3145P}. This order was
selected due to several strong line of thorium in this region. The 
calibration of the Lick spectrum was performed using only \ion{Ar}{1} and
\ion{Ar}{2} lines. The spectra demonstrate apparent dissimilarities. Only argon
lines are common for both spectra, whereas numerous thorium lines are absent in
the Lick spectrum. Inset of Fig.\ref{fig:cmp1} demonstrates this point in more
details. Yet the Lick spectrum contains a huge number of strong unidentified
lines which are not observed in the NTT spectrum. In this regard we should
consider a possibility that the lamp cathode is made of another material
than the thorium.

In attempt to resolve the issue of unknown spectral lines, we consider the
114-th order (Fig.~\ref{fig:cmp2}), in which case we find five lines of
\ion{Ar}{2} to start the wavelength calibration and 15 unidentified lines with
the relative intensity $>0.1$. The accuracy of the preliminary calibration is
$\sigma\approx0.002$~\AA\ using a polynomial approximation of the third order.
Positions of all unidentified lines match neither thorium nor neon
likewise in case of the 110-th order. The unknown lines obviously should belong
to some other element. To define this element we compiled from the VALD
database \citep{1999A&AS..138..119K} two line lists: the first contains lines
that coincide with wavelengths of unknown lines with accuracy
$\Delta\lambda=\pm0.03$~\AA, while the second with accuracy
$\Delta\lambda=\pm0.01$~\AA. If unknown lines belong to one element then
it should meet at least 15 times in the list. For case two elements, the
sum of its lines should be 15. When we reduce accuracy from 0.03 to
0.01~\AA\ the required element or elements should remain in the list, while
others elements should be separated. The first line list reveals 16 lines of
\ion{Ti}{1} and 10 lines of \ion{Fe}{2}; for the rest of elements the number of
lines $<5$. The second list contains 14 lines of \ion{Ti}{1} and 4 lines of
\ion{Fe}{2}; for others elements the number of lines $<2$. Thus, titanium and
iron are possible elements that form the unknown lines in the lamp spectrum.

\begin{deluxetable}{rrrrrr} 
\tablecaption{List of characteristics of \ion{Ti}{1} lines related with unknown
lines from the 114-th echelle order \label{tab:Ti114} }
\tablecolumns{6}
\tablewidth{0pt}
\small
\tablehead{
  \colhead{$\lambda_{meas}$} &
  \colhead{ $\lambda_{Ti\,I}$} &
  \colhead{$\Delta\lambda$} &
  \colhead{$E$} &
  \colhead{log\,$gf$}& 
  \colhead{$I_{rel}$}
\\
  \colhead{\AA}&
  \colhead{\AA}&
  \colhead{\AA}&
  \colhead{eV} &
  &
}
\startdata
4981.727 & 4981.731 &  0.004 & 3.337 &  0.577 & 0.091 \\
4991.063 & 4991.065 &  0.002 & 3.319 &  0.467 & 0.091 \\
4997.093 & 4997.096 &  0.003 & 2.480 & -2.070 & 0.021 \\
4999.499 & 4999.503 &  0.004 & 3.305 &  0.359 & 0.095 \\
5007.206 & 5007.210 &  0.004 & 3.294 &  0.258 & 0.083 \\
5014.203 & 5014.276 &  0.073 & 3.285 &  0.332 & 0.141 \\
5016.158 & 5016.161 &  0.003 & 3.319 & -0.574 & 0.012 \\
5020.023 & 5020.026 &  0.003 & 3.305 & -0.414 & 0.020 \\
5022.864 & 5022.868 &  0.004 & 3.294 & -0.434 & 0.024 \\
5024.843 & 5024.844 &  0.001 & 3.285 & -0.602 & 0.016 \\
5035.902 & 5035.903 &  0.001 & 3.921 &  0.492 & 0.011 \\
5036.463 & 5036.464 &  0.001 & 3.904 &  0.239 & 0.014 \\
5038.397 & 5038.397 &  0.000 & 3.890 &  0.159 & 0.012 \\
5039.957 & 5039.957 &  0.000 & 2.480 & -1.236 & 0.203 \\
5064.644 & 5064.653 &  0.009 & 2.495 & -1.059 & 0.161 \\
\enddata
\end{deluxetable}

First, we test the iron because it basically can be used for the lamp 
cathode as well. In an astronomical practice FeAr lamp is used sometimes for the
wavelength calibration. For example, at the Terskol observatory of Institute of
Astronomy of Russian Academy of Sciences the wavelength calibration is performed
with the FeAr lamp \citep{2009ARep...53..685P}. The line list for the FeAr lamp
can be found in MIDAS package. We checked that unknown lines might belong to
iron and ruled out this possibility due to unsuccessfully calibration using the
FeAr list and absence of the strongest iron lines in the spectrum of the studied
lamp.

We therefore turned to the titanium as the likely source of strong and weak
lines seen in the 114-th order (Tabl.~\ref{tab:Ti114}). The table contains
measured wavelength, laboratory wavelength of \ion{Ti}{1}, their difference, the
energy of upper level $E$, the oscillator strength $\log\,gf$, and the relative
line intensity $I_{rel}$. The wavelength coincidence is remarkable with the
exception of 5014.276~\AA\ line in which case $\Delta\lambda$=0.073\AA. This
line, however, is blended with another line of \ion{Ti}{1} at
$\lambda$=5014.187\AA\ ($E$=2.472~eV and $\log\,gf$=-1.462). It is noteworthy,
all the identified \ion{Ti}{1} lines have large transition probabilities and low
excitation potential, both properties favouring relatively high line intensity
in accordance with the observed lamp spectrum. This rule holds for the other
echelle orders as well. We therefore conclude that the Lick lamp cathode is
made of titanium. 

This conclusion seems rather unexpected given the fact that the lamp is
designated as ThAr. On the other hand, hallow-cathode lamps with Ti cathode are
widely manufactured and normally used for spectral calibration in physics and 
chemistry. The company S\&J Juniper \& Co also has TiAr lamp in products
(catalog number is 4163). We conjecture that the Ti cathode lamp has got into
the ThAr lamp shipment by mistake. 

In the astronomical practice Ti cathode lamps are not used which explains why Ti
line list is absent in MIDAS and other packages designed for processing of
echelle spectra. Therefore, it is important to create a list of spectral lines
of titanium in optical range to use this hollow-cathode lamp installed at the
Hamilton spectrograph.

\section{Titanium line list}

The line list of \ion{Ti}{1}, which would corresponds to observed special
lines, can be retrieved from the VALD database. To this
end we have to estimate the temperature of the line-emitting gas using measured
line intensities. We employ standard approach based on the assumption of the
Boltzmann level population and neglecting self-absorption effects. In this case,
the line flux $F\propto (gf/\lambda^3)\exp(-E/kT)$, where $g$ is the statistical
weight of the upper level, $f$ is the oscillator strength for the transition,
$E$ is the excitation energy of the upper level, $k$ is the Boltzmann constant,
and $T$ is the excitation temperature.  

In the logarithmic form we come to the linear dependence on the excitation 
potential (aka Boltzmann plot)
\begin{equation}
\log\frac{gf}{F\lambda^3} = a+bE,
\label{eq:lgI}
\end{equation}
where $a$ is a constant which determines the absolute scale of oscillator
strengths, and $b=5040/T$ is a constant representing the decrement of level
population.

\begin{figure}[t]
\centering
\resizebox{\hsize}{!}{\includegraphics[clip]{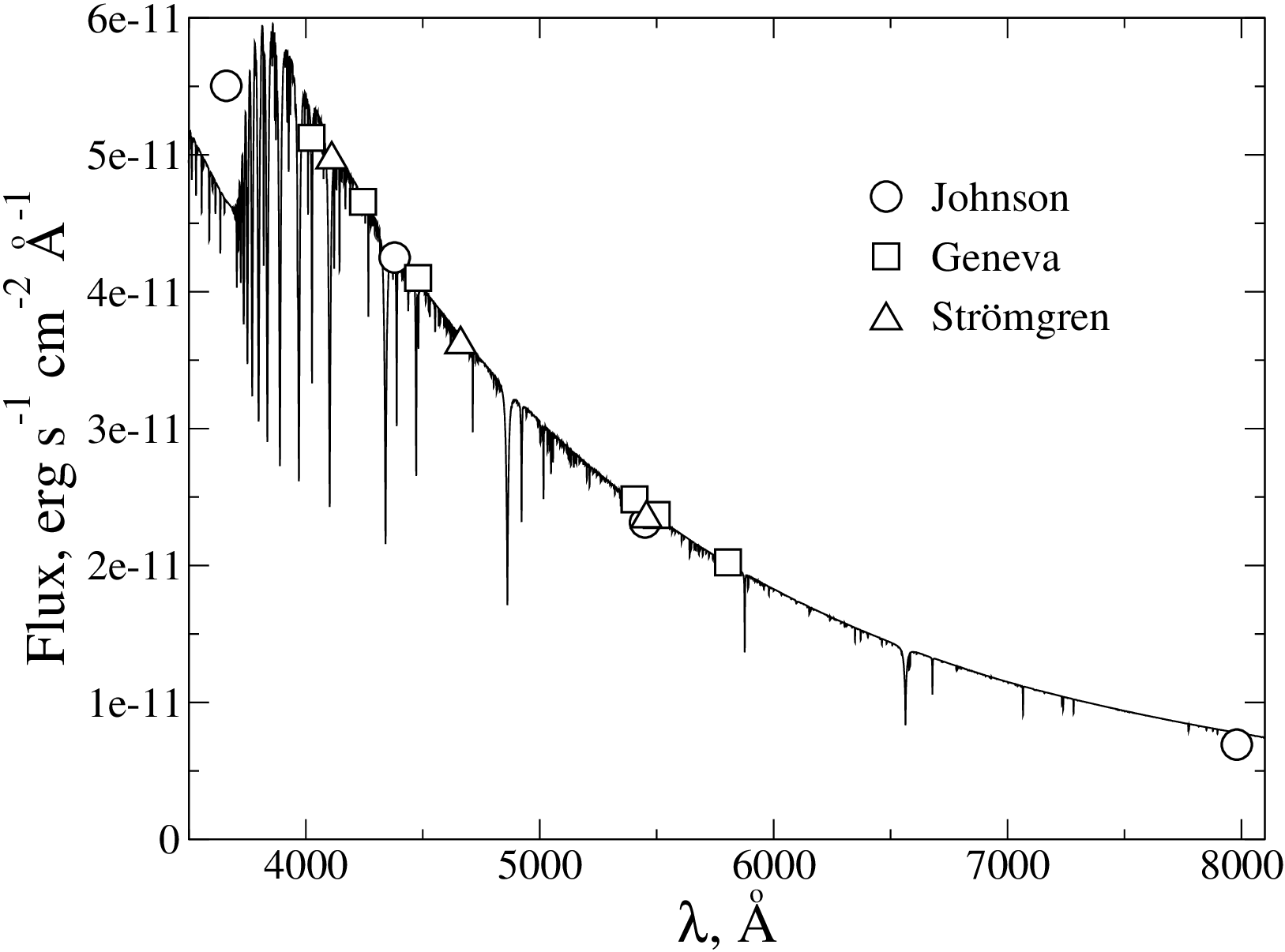}}
\figcaption{Theoretical spectral energy distribution of the HR 1215 radiation flux 
and the flux recovered from photometric data shown by different symbols. 
The error bars of photometry fluxes are well within symbol size.}
\label{fig:HR1215}
\end{figure}

\subsection{Flux calibration}

The temperature, i.e., the constant $b$ of Eq.~\ref{eq:lgI}, can be determined,
if one knows line calibrated intensities. To recover line intensities from the
echelle spectra one needs the blaze function and energy distribution for the
echelle orders. The absolute calibration was performed using the echelle spectra
of B-star HR~1215 taken in the same observations. Since this star is not
spectrophotometric standard we recover its spectral energy distribution (SED)
via the determination of the stellar effective temperature \Teff\ and gravity
\lgg. The LSR star velocity components $(U,V,W)$=($18.9$, $-0.8$,
$-4.8$)~km\,s$^{-1}$ are within the velocity dispersion of the thin disk in the
solar neighbourhood \citep{2005A&A...430..165F}. We assume, therefore, the
solar metallicity. 

Parameters of the stellar atmosphere are estimated via a two-step procedure. 
First, using photometric data and calculating the preliminary SED we determine
the initial blaze function of echelle and extract the stellar spectrum. On the
second step we fit the observed spectrum by the synthetic stellar spectrum and
calculate the final blaze function. 

The first step. The preliminary values of \Teff\ and reddening
\mbox{${E(B-V)}$} are found from the two color diagram
\mbox{$(U-B)$}-\mbox{$(B-V)$}. The observed color indices are
\mbox{$(B-V)$}=-0.03 and \mbox{$(U-B)$}=-0.74 (cf. SIMBAD and
\citep{1960IzKry..22..257B}). These colors combined with the curve of the colors
of normal stars results in the reddening estimate ${E(B-V)}=0.20$~mag. The
deredenned color is then \mbox{$(B-V)_0$}=-0.23 which corresponds to the
effective temperature \Teff$\approx$24000~K. The gravity is estimated using the
standard relation:
\begin{eqnarray}\nonumber
\log\,g=-10.607+\log\,(M/M_\odot)+4\log\,T_{eff}-\\ \nonumber
-0.4(4.69-V-5-5\log\,\pi-A_v-BC), 
\end{eqnarray}
where $M/M_\odot$ is the stellar mass in solar units, $V=5.49$~mag,
$A_V=3.1E(B-V)$ are magnitude and absorption in V-band, $BC$ is the bolometric
correction, $\pi$=3.76$\pm$0.35 is the Hipparcos parallax
\citep{2007A&A...474..653V}. With the mass $M$=9$\pm1~M_\odot$ estimated from
evolutionary tracks \citep{1992A&AS...96..269S} one obtains \lgg=4.0$\pm$0.1.
With the derived values of \lgg\ and \Teff\  we compute the stellar atmosphere 
model using \emph{ATLAS9} code \citep{1993KurCD..13.....K}. The synthetic
spectrum is then calculated using \emph{SynthVb} code
\citep{2003IAUS..210P.E49T}. This spectrum is converted into the MIDAS image
format to construct the initial blaze function.

The second step. The extracted spectrum of HR~1215 with the preliminary
absolute flux calibration is used to estimate more reliable stellar parameters
via stellar spectrum synthesis. The effective temperature \Teff\ and gravity
\lgg\ are found by the Balmer lines fit (cf. \citep{2010A&A...519A..86I}). The
derived parameter values are \Teff=20000$\pm$1000~K and \lgg=3.9$\pm$0.1. The
corresponding normal color \mbox{$(B-V)$}=-0.20~mag, the reddening
\mbox{$E(B-V)$}=0.17~mag and $A_v=0.5$~mag. For these parameters the model of
the stellar atmosphere and synthetic spectrum are calculated with the reddening
taken into account according to extinction data of \citet{1990ARA&A..28...37M}
and code REDDEN \citep{1993KurCD..13.....K}. 

The synthesised SED of HR~1215 can be calibrated to produce the absolute flux
using photometric data for Johnson \citep{1998A&A...333..231B}, Geneva
\citep{1988A&A...206..357R}, and Str\"{o}mgren \citep{1998AJ....116..482G} 
photometric systems. The normalized synthesised stellar SED turns out to be 
in excellent agreement with the SED calculated from photometric data using
\emph{SynthVb} (Fig.~\ref{fig:HR1215}). Finally, the absolute SED of HR~1215 is
used to produce the final calibrated blaze function.

\begin{figure}[t]
\centering
\resizebox{\hsize}{!}{\includegraphics[clip]{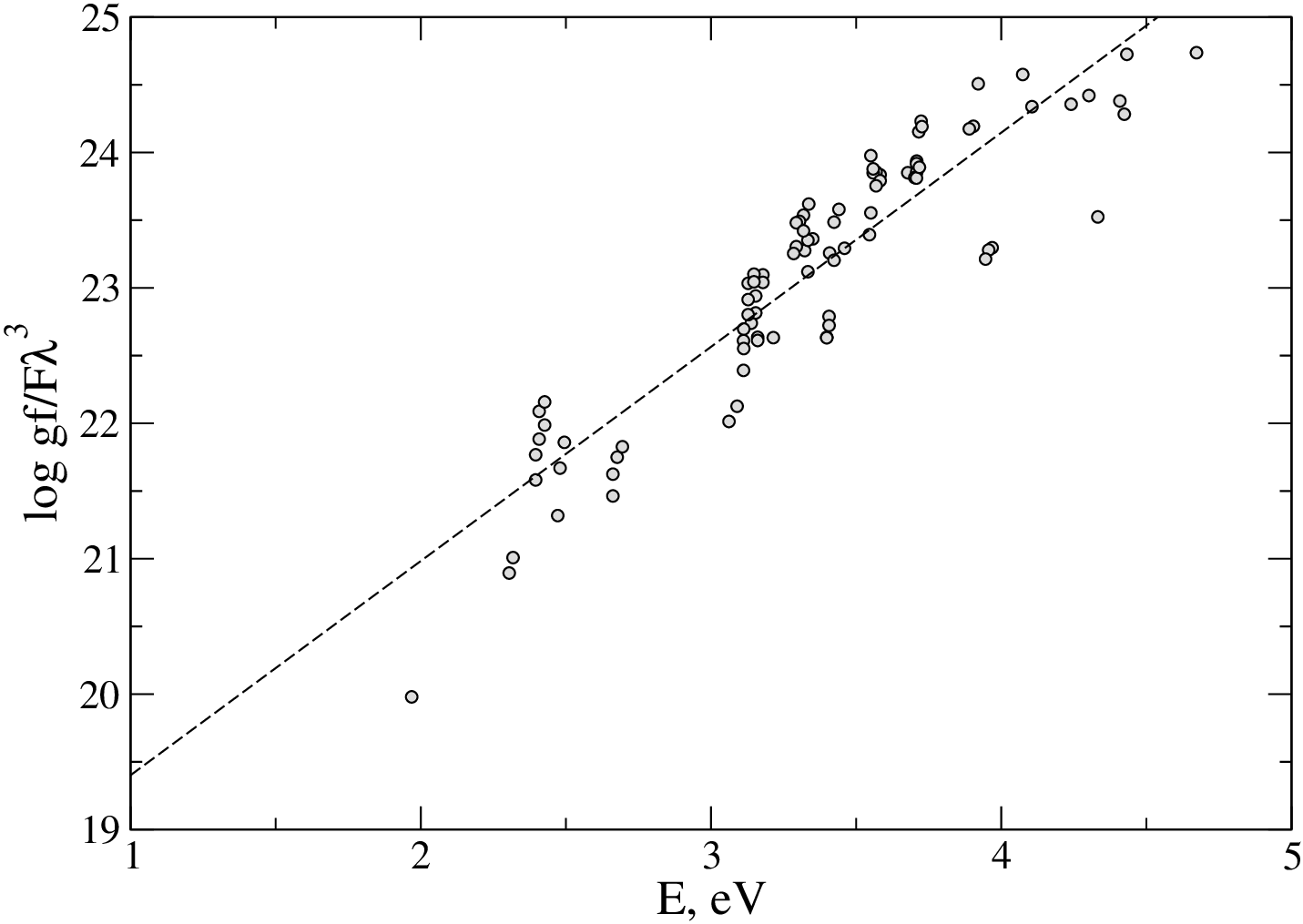}}
\figcaption{Estimation of the temperature of \ion{Ti}{1} atoms in the
hallow-cathode lamp
by eq.~\ref{eq:lgI}}
\label{fig:I-E}
\end{figure}

\subsection{Excitation temperature of \ion{Ti}{1}}

The observed lamp spectrum of \ion{Ti}{1} is reduced by the final blaze
function to produce the spectrum in relative intensities. All the observed
lines of \ion{Ti}{1} in the whole spectral range have been identified by the
VALD database. Their relative intensities are used to infer temperature from the
Boltzmann plot (Fig.~\ref{fig:I-E}). A typical error in $\log\,(gf/F\lambda^3)$
is about 0.2-0.5~dex; the uncertainty originates from the spectrum noise, errors
of the blaze function, and errors of oscillator strengths. The error is larger
for weak lines, i.e., for larger $E$. From the linear least-square fit we obtain
$a=17.82\pm0.25$ and $b=1.582\pm0.076$ (Eq.~\ref{eq:lgI}); the latter suggests 
the excitation temperature of $5040/b=3200\pm150$~K.

\begin{table}
\caption{Final \ion{Ti}{1} and \ion{Ti}{2} line list 
Table is published in its entirety in the
electronic edition. A portion is shown here for guidance regarding its 
form and content. \label{tab:linelist}}
\centering
\begin{tabular}{rrl}
\tableline
 \header{Wavelength} & \header{$I_{rel}$} &  \header{Ion}\\
 \header{\AA}        &           &     \\
\tableline
 3000.865 &   20.4 & \ion{Ti}{1}\\
 3002.725 &    1.4 & \ion{Ti}{1}\\
 3042.540 &    1.7 & \ion{Ti}{1}\\
 3088.025 &  715.2 & \ion{Ti}{2}\\
 3100.663 &    1.3 & \ion{Ti}{1}\\
 3119.723 &    5.0 & \ion{Ti}{1}\\
\tableline
\end{tabular}
\end{table}

\subsection{Titanium line list}

We find that the detection limit for center of the \ion{Ti}{1} line flux in the
lamp spectrum is
$F_{\lambda,min}\approx10^{-12}$~erg\,s$^{-1}$\,cm$^{-2}$\,\AA$^{-1}$ with the
integrated flux $F_{min} =1.77\,F_{\lambda,min}\lambda/R$, where $R=60000$ is
the resolving power. We, therefore, extract from the VALD all the \ion{Ti}{1}
lines in the range of 3000-12000~\AA\ with the fluxes
$F>F_{min}=1.77\,F_{\lambda,min}\lambda/R$,
or
$$
F_{\lambda}=\frac{R~gf}{1.77 \lambda^4} 10^{-17.82-1.582 E}>10^{-12}
$$

\begin{figure}[t]
\centering
\resizebox{\hsize}{!}{\includegraphics[clip]{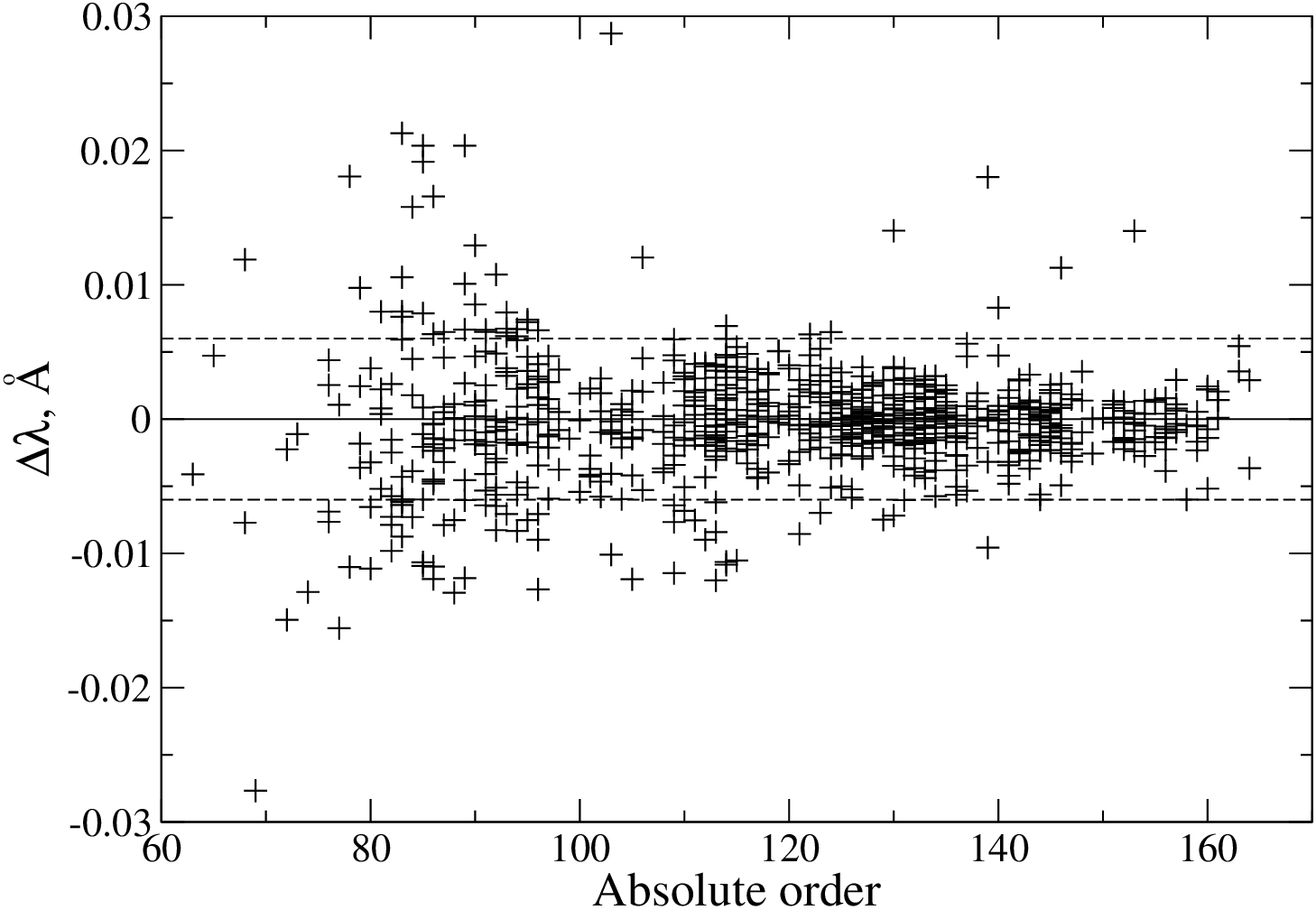}}
\figcaption{The residual wavelengths for final calibration of the Hamilton
echelle spectrograph}
\label{fig:calib}
\end{figure}

The number of \ion{Ti}{1} lines that meets this requirement is 585, a small
fraction of complete VALD \ion{Ti}{1} line list (16110 lines). About 80\% of
observed lines fall within the range of 3000-7000~\AA. 

We find six lines of \ion{Ti}{2} in the ultraviolet region ($\lambda<3761$\AA)
of the observable spectrum. Using measured intensities of \ion{Ti}{1} and
\ion{Ti}{2} lines and assuming the Boltzmann level population we derive the
number ratio of ions $N$(\ion{Ti}{2})/$N$(\ion{Ti}{1})$\approx0.25$, which
indicates rather high ionization fraction of \ion{Ti}{2} ($\sim 0.2$). 
\ion{Ti}{2} is presented by 14 lines in our line list. The blending effect is
not taken into account in the final list of \ion{Ti}{1} and \ion{Ti}{2} lines.
For the resolving power $R=60000$ the line list contains 28 blended lines with
the separation less than 0.1\AA. The final list of titanium lines is presented
in online Table~\ref{tab:linelist}, where the first column is wavelength, the
second is calculated relative intensity, and the last column is titanium ion.

\section{Wavelength calibration of the Hamilton spectrograph}

The TiAr spectrum taken on the Hamilton spectrograph is calibrated using the
line list produced by merging of \ion{Ti}{1} and \ion{Ti}{2} line list with the
Ar line list from NOAO \citep{NOAO_ThAr}. The first extracted order (absolute
number is 170) and the last order (absolute number is 56) have spectral ranges
3315.71 -- 3402.41~\AA\ and  10063.24 -- 10326.02~\AA\ respectively. Wavelength
residuals of the calibration are ploted in Fig.~\ref{fig:calib}. The average
deviation is $\Delta \lambda\approx0.006$~\AA.
The accuracy is less than 0.01~\AA\ for orders above 90 ($\lambda\lesssim6500$),
while in the red the average accuracy is about 0.01~\AA. The accuracy 
deteriorates towards the red region because the strong Ar lines produce light
pollution on the CCD, and because the red part contains too few reference lines.
In this region the calibration was made by solving the standard echelle grating
equation. 

\section{SUMMARY}

Thus, in this study we show that in spectrum of used hallow-cathode ''TiAr'' 
lamp the lines of thorium are absent, whereas numerous unknown spectral 
lines are presented. Due to this fact, the calibration of the Hamilton
spectrograph in all echelle orders was very difficult or even impossible. To
solve the problem we state that unknown lines  belong to titanium. We estimate
the temperature of gas in the lamp, and create the theoretical list of the
relevant \ion{Ti}{1} and \ion{Ti}{2} lines that should be observed in
the spectra. Using this line list compiled with list of argon lines the
calibration of the Hamilton spectrograph was made with accuracy better than
0.01~\AA.

\acknowledgments

The authors thank Chugai~N.~N. for a discussion and comments. 
The paper essentially used results of the project VAMDC of European Commission
(No.~239108). This work was supported by the "Nonstationary Phenomena in Objects
of the Universe" Program of the Presidium of the Russian Academy of Sciences,
the Federal Agency for Science and Innovations (grant No.~8529), and the
National Natural Science Foundation of China (grant No.~11243004).

Facilities: \facility{Shane (Hamilton spectrograph)}

\bibliographystyle{abbrvnat}
\bibliography{paper}

\end{document}